\documentstyle[epsf]{mn}

\newif\ifAMStwofonts

\newcommand{\chisq}{\hbox{$\chi^2$}}
\newcommand{\bj}{\hbox{$b_{J}$}}
\newcommand{\fabs}{\hbox{$f_{\rm Lyc}^{abs}$}}
\newcommand{\hi}{\hbox{H\,{\sc i}}}
\newcommand{\hii}{\hbox{H\,{\sc ii}}}

\newcommand{\oh}{\hbox{O/H}}

\newcommand{\ha}{\hbox{H$\alpha$}}

\newcommand{\hb}{\hbox{H$\beta$}}

\newcommand{\oii}{\hbox{[O\,{\sc ii}]}}
\newcommand{\oiii}{\hbox{[O\,{\sc iii}]}}
\newcommand{\nii}{\hbox{[N\,{\sc ii}]}}
\newcommand{\sii}{\hbox{[S\,{\sc ii}]}}

\newcommand{\hahb}{\hbox{H$\alpha$/H$\beta$}}

\newcommand{\oiiha}{\hbox{[O\,{\sc ii}]/H$\alpha$}}

\newcommand{\oiiihb}{\hbox{[O\,{\sc iii}]/H$\beta$}}
\newcommand{\oiioiii}{\hbox{[O\,{\sc ii}]/[O\,{\sc iii}]}}

\newcommand{\siiha}{\hbox{[S\,{\sc ii}]/H$\alpha$}}
\newcommand{\niiha}{\hbox{[N\,{\sc ii}]/H$\alpha$}}
\newcommand{\niisii}{\hbox{[N\,{\sc ii}]/[S\,{\sc ii}]}}
\newcommand{\oiisii}{\hbox{[O\,{\sc ii}]/[S\,{\sc ii}]}}

\newcommand{\htauv}{\hbox{$\hat{\tau}_V$}}
\newcommand{\htauvhii}{\hbox{$\hat{\tau}_V^{\rm HII}$}}
\newcommand{\htauvhi}{\hbox{$\hat{\tau}_V^{\rm HI}$}}
\newcommand{\htauvism}{\hbox{$\hat{\tau}_V^{\rm ISM}$}}

\newcommand{\abswha}{\hbox{$W_{{\rm H}\alpha}^{\rm abs}$}}
\newcommand{\abswhb}{\hbox{$W_{{\rm H}\beta}^{\rm abs}$}}

\newcommand{\lamha}{\hbox{$\lambda_{{\rm H}\alpha}$}}
\newcommand{\Aha}{\hbox{$A_{{\rm H}\alpha}$}}
\newcommand{\Av}{\hbox{$A_{V}$}}
\newcommand{\Ahamed}{\hbox{$\bar{A}_{{\rm H}\alpha}$}}
\newcommand{\Avmed}{\hbox{$\bar{A}_{V}$}}

\newcommand{\lstar}{\hbox{$L_{\ast}$}}
\newcommand{\lha}{\hbox{$L_{{\rm H}\alpha}$}}
\newcommand{\sfr}{\hbox{${\rm SFR}$}}
\newcommand{\sfrha}{\hbox{${\rm SFR}_{{\rm H}\alpha}$}}
\newcommand{\sfroii}{\hbox{${\rm SFR}_{\rm [O\,{\sc ii}]}$}}
\newcommand{\sfrfir}{\hbox{${\rm SFR}_{\rm FIR}$}}
\newcommand{\sfruv}{\hbox{${\rm SFR}_{\rm UV}$}}

\newcommand{\loii}{\hbox{$L_{\rm [O\,{\sc ii}]}$}}

\newcommand{\lfir}{\hbox{$L_{\rm FIR}$}}
\newcommand{\luvnu}{\hbox{$L_{\nu,{\rm UV}}$}}

\newcommand{\uavo}{\hbox{$\langle {\tilde{U}}_0\rangle$}}

\newcommand{\nav}{\hbox{$\tilde{n}_{\rm H}$}}
\newcommand{\xav}{\hbox{$\tilde{\xi}_{\rm d}$}}
\newcommand{\zav}{\hbox{$\tilde{Z}$}}

\ifoldfss
  \ifCUPmtlplainloaded \else
    \NewTextAlphabet{textbfit} {cmbxti10} {}
    \NewTextAlphabet{textbfss} {cmssbx10} {}
    \NewMathAlphabet{mathbfit} {cmbxti10} {} 
    \NewMathAlphabet{mathbfss} {cmssbx10} {} 
  \fi
  \ifAMStwofonts
    \ifCUPmtlplainloaded \else
      \NewSymbolFont{upmath} {eurm10}
      \NewSymbolFont{AMSa} {msam10}
      \NewMathSymbol{\upi}     {0}{upmath}{19}
      \NewMathSymbol{\umu}     {0}{upmath}{16}
      \NewMathSymbol{\upartial}{0}{upmath}{40}
      \NewMathSymbol{\leqslant}{3}{AMSa}{36}
      \NewMathSymbol{\geqslant}{3}{AMSa}{3E}

      \let\leq=\leqslant 
       
    \fi
  \fi
\fi 

\ifnfssone
  \newmathalphabet{\mathit}
  \addtoversion{normal}{\mathit}{cmr}{m}{it}
  \addtoversion{bold}{\mathit}{cmr}{bx}{it}
  \newmathalphabet{\mathbfit} 
  \addtoversion{normal}{\mathbfit}{cmr}{bx}{it}
  \addtoversion{bold}{\mathbfit}{cmr}{bx}{it}
  \newmathalphabet{\mathbfss} 
  \addtoversion{normal}{\mathbfss}{cmss}{bx}{n}
  \addtoversion{bold}{\mathbfss}{cmss}{bx}{n}
  \ifAMStwofonts
    \ifCUPmtlplainloaded \else
      \UseAMStwoboldmath
      \makeatletter
      \new@mathgroup\upmath@group
      \define@mathgroup\mv@normal\upmath@group{eur}{m}{n}
      \define@mathgroup\mv@bold\upmath@group{eur}{b}{n}
      \edef\UPM{\hexnumber\upmath@group}
      \new@mathgroup\amsa@group
      \define@mathgroup\mv@normal\amsa@group{msa}{m}{n}
      \define@mathgroup\mv@bold\amsa@group{msa}{m}{n}
      \edef\AMSa{\hexnumber\amsa@group}
      \makeatother
      \mathchardef\upi="0\UPM19
      \mathchardef\umu="0\UPM16
      \mathchardef\upartial="0\UPM40
      \mathchardef\leqslant="3\AMSa36
      \mathchardef\geqslant="3\AMSa3E

      \let\leq=\leqslant 

    \fi
  \fi
\fi 

\ifnfsstwo
  \DeclareMathAlphabet{\mathbfit}{OT1}{cmr}{bx}{it}
  \SetMathAlphabet\mathbfit{bold}{OT1}{cmr}{bx}{it}
  \DeclareMathAlphabet{\mathbfss}{OT1}{cmss}{bx}{n}
  \SetMathAlphabet\mathbfss{bold}{OT1}{cmss}{bx}{n}
  \ifAMStwofonts
    \ifCUPmtlplainloaded \else
      \DeclareSymbolFont{UPM}{U}{eur}{m}{n}
      \SetSymbolFont{UPM}{bold}{U}{eur}{b}{n}
      \DeclareSymbolFont{AMSa}{U}{msa}{m}{n}
      \DeclareMathSymbol{\upi}{0}{UPM}{"19}
      \DeclareMathSymbol{\umu}{0}{UPM}{"16}
      \DeclareMathSymbol{\upartial}{0}{UPM}{"40}
      \DeclareMathSymbol{\leqslant}{3}{AMSa}{"36}
      \DeclareMathSymbol{\geqslant}{3}{AMSa}{"3E}

      \let\leq=\leqslant 

    \fi
  \fi
\fi 

\ifCUPmtlplainloaded \else
  \ifAMStwofonts \else 
    \def\upi{\pi}
    \def\umu{\mu}
    \def\upartial{\partial}
  \fi
\fi

\title{Star formation, metallicity and dust properties derived from the 
SAPM galaxy survey spectra}
\author[S. Charlot et al.]{
\parbox[t]{\textwidth}{
S.~Charlot$^{1,2}$\footnotemark,
G.~Kauffmann$^{2}$,
M.~Longhetti$^{3}$,
L.~Tresse$^{4}$,
S.~D.~M.~White$^{2}$,
S.~J.~Maddox$^{5}$
and S.~M.~Fall$^{6}$}
\vspace*{6pt} \\
$^1$Institut d'Astrophysique de Paris, CNRS, 98 bis Boulevard Arago, 75014 Paris, France\\
$^2$Max-Planck Institut f\"ur Astrophysik, Karl-Schwarzschild-Strasse 1, 85748 Garching, Germany\\
$^3$Osservatorio Astronomico di Brera, Via Bianchi 46, 23807 Merate (LC), Italy\\
$^4$Laboratoire d'Astrophysique de Marseille, CNRS, Les Trois-Lucs,
B.P. 8, 13376 Marseille Cedex 12, France\\
$^5$School of Physics \& Astronomy, University of Nottingham, Nottingham NG7 2RD, UK\\
$^6$Space Telescope Science Institute, 3700 San Martin Drive, Baltimore, MD 21218, USA\\
}

\date{Accepted for publication in MNRAS}

\pagerange{\pageref{firstpage}--\pageref{lastpage}}
\pubyear{1994}

\begin{document}

\maketitle

\label{firstpage}

\begin{abstract}

We have derived star formation rates (SFRs), gas-phase oxygen abundances and effective
dust absorption optical depths for a sample of galaxies drawn from the Stromlo-APM 
redshift survey using the new Charlot \& Longhetti (2001; CL01) models, which provide
a physically consistent description of the effects of stars, gas and dust on the 
integrated spectra of galaxies.  Our sample consists of 705 galaxies with measurements
of the fluxes and equivalent widths of \ha, \oii, and one or both of \nii\ and \sii. For 
a subset of the galaxies, 60 and 100 $\mu$m IRAS fluxes are available. We compare the star
formation rates derived using the models with those derived using standard estimators
based on the \ha, the \oii\ and the far-infrared luminosities of the galaxies. The CL01
SFR estimates agree well with those derived from the IRAS fluxes, but are typically a 
factor of $\sim3$ higher than those derived from the \ha\ or the \oii\ fluxes, even 
after the usual mean attenuation correction of $\Aha=1$ mag is applied to the data. We
show that the reason for this discrepancy is that the standard \ha\ estimator neglects 
the absorption of ionizing photons by dust in \hii\ regions and the contamination 
of \ha\ emission by stellar absorption. We also use our sample to study variations in
star formation and metallicity as a function of galaxy absolute \bj\ magnitude. For this
sample, the star formation rate per unit \bj\ luminosity is independent of magnitude.
The gas-phase oxygen abundance does increase with \bj\ luminosity,
although the scatter in metallicity at fixed magnitude is large.

\end{abstract}


\section{Introduction}

\footnotetext{Email: charlot@iap.fr}

Most large spectroscopic galaxy surveys have been carried out with the aim
of determining the luminosity function or of studying the three-dimensional
clustering properties of galaxies. The spectra are used to measure 
redshifts and so to infer the distances to galaxies.   
However, the spectra also contain a wealth of additional information about the
physical properties of galaxies. The continuum and the absorption lines provide
information about the stellar content, while the nebular emission lines provide a 
measure of the star formation rate (SFR) and the interstellar metallicity.                

Kennicutt (1992a) published an atlas of high-resolution, high signal-to-noise,
flux-calibrated spectra of a small sample of nearby galaxies of different Hubble 
types. The spectra obtained in redshift surveys are typically of much 
lower quality than the spectra in the Kennicutt atlas, but
they nevertheless provide a means of studying the star formation rates
and metallicities of a magnitude-limited sample of
normal galaxies and of characterizing trends in these properties as
a function of galaxy luminosity and environment. As a result,
there have been a large number of papers analyzing the spectral properties
of redshift survey galaxies. 

Many studies focus on SFR estimates based on a single strong emission 
line, such as \oii$\lambda$3727 (e.g. Hashimoto et al 1998; Christlein
2000; Baldi, Bardelli \& Zucca 2001). One of the advantages
of \oii\ is that it is easily observed in high-redshift galaxies and can
thus be used to study evolutionary trends in star formation (e.g.
Ellis et al 1996; Cowie et al 1996;  Hammer et al 1997; Balogh et al 1998).
Other studies focus on additional lines, in particular
the H$\alpha$ line, which is believed to be a more robust tracer of
star formation than \oii\ (e.g. Gallego et al 1995; Tresse \& Maddox 1998; 
Tresse et al 1999; Barton, Geller \& Kenyon 2000).

Kennicutt (1998; hereafter K98) reviewed a number of traditional
diagnostic methods for inferring star formation rates from galaxy spectra.
The two methods that are applicable to redshift survey data
are based on the H$\alpha$ recombination line and the \oii\
forbidden-line doublet. As discussed by K98, the derived SFRs are subject
to errors because of attenuation by dust. In the case of \oii, variations
in excitation, metallicity and diffuse gas fraction will also contribute 
to the uncertainty in the derived SFRs. Both methods ignore the absorption
of ionizing photons by dust in \hii\ regions, which could play an important
role (Petrosian, Silk \& Field 1972; Mathis 1986; Charlot \& Fall 2000).

Recently, Charlot \& Longhetti (2001, hereafter CL01) {\em quantified} the 
uncertainties in these SFR estimators. They constructed physically realistic
models describing the propagation of the photons emitted by stars through the
interstellar medium (ISM) and their absorption by dust in galaxies. These models
combine the latest version of the Bruzual \& Charlot (1993) population synthesis
code with the Ferland (1996) photoionization code for modelling nebular emission.
Stars are assumed to form in interstellar birth clouds (i.e. giant molecular 
clouds). After $10^7\,$yr, young stars disrupt their birth clouds and migrate
into the ambient ISM. The line emission from a galaxy arises both from the \hii\
regions surrounding young stars and from the diffuse gas ionized by photons
that have leaked into the ambient ISM. In the CL01 model, the depletion
of heavy elements onto dust grains, the absorption of ionizing photons by dust,
and the contamination of Balmer emission lines by stellar absorption are all
included in a self-consistent way. The absorption of photons from \hii\ regions
and from older stars is described by a dust model developed by Charlot \& Fall
(2000). In this prescription, line and continuum photons are attenuated
differently because of the finite lifetimes of the clouds in which stars form.

CL01 showed that in models that include absorption by dust and that match the
observed ionized-gas properties of local galaxies, SFR estimates based
purely on the H$\alpha$ or \oii\ luminosities of galaxies can be in error by more
than an order of magnitude. On the other hand, with the help of other lines such as
H$\beta$, \oiii, \nii\ and \sii, these errors can be substantially reduced.
CL01 presented a number of new estimators of the SFRs, gas-phase oxygen
abundances and dust absorption optical depths of galaxies based on combinations
of these lines.

This paper presents the first application of the CL01 model to a magnitude-limited
sample of galaxies drawn from the Stromlo-APM (SAPM) redshift survey (Loveday et al
1996). Tresse et al (1999) have measured the fluxes and equivalent widths of
\ha , \oii, \nii\ and \sii\ for the galaxies in this survey. We compare the 
SFRs derived using standard estimators based on the H$\alpha$ and \oii\ lines
to those derived using the CL01 models, which make use of all four lines. We show 
that the CL01 SFR estimates are typically a factor of $\sim3$ higher than those
derived using just \ha\ or \oii, {\em even after} the usual mean attenuation
correction of $\Aha=1$ mag is applied to the data. We show that the reason for
this discrepancy is that the standard \ha\ estimator neglects the absorption of
ionizing photons by dust in \hii\ regions and the contamination of \ha\ emission 
by stellar absorption.

Observations from the {\it Infrared Astronomical Satellite} (IRAS) are available for
149 galaxies in the sample and provide an independent check on the derived star
formation rates. We find that the CL01 SFR estimates agree surprisingly well with
those based on the 60 and 100 $\mu$m IRAS fluxes.

Within the subset of the Stromlo-APM survey which we study, we find that the
star formation rate per unit $b_J$ luminosity is independent of galaxy absolute
magnitude. We derive gas-phase oxygen abundances and dust absorption optical
depths for this sample.  Bright, \lstar\ galaxies have O/H values close
to solar, while fainter galaxies are found to be less metal-rich on average. The
distribution of \ha\ attenuations is similar to that found for other samples of
nearby star-forming galaxies.
 
We describe the properties of the SAPM galaxy sample in Section 2 below. In
Section 3, we illustrate the inconsistencies obtained when estimating star 
formation rates of galaxies using different traditional diagnostic methods
(\ha, \oii, UV and FIR luminosities).  Readers familiar with the uncertainties
affecting star formation measures in galaxies may skip at least the first part
of this section. In Section 4, we describe how the new models developed by CL01
can help us better constrain the star formation, metallicity and dust properties
of SAPM galaxies using a combination of different spectral diagnostics. In 
Section 5, we present the results of this analysis and identify the important
weaknesses affecting the traditional \ha\ estimator. Our conclusions are 
summarized in Section 6.

\section{Description of the sample}

Our sample is drawn from the                                              
the Stromlo-APM (SAPM) redshift survey (Loveday et al. 1996). The 
SAPM survey consists of 1797 galaxies brighter than $\bj=17.15$ 
covering 4300~deg$^2$ of the south galactic cap. The mean redshift of  
the galaxies in the survey is                                                
$\langle z \rangle=0.051$. The spectra were taken with the 
Dual-Beam Spectrograph of the Australian National University 2.3-m 
telescope at Siding Spring.  The wavelength range is 3700--5000~{\AA} in
the blue and 6300--7600~{\AA} in the red, and the dispersion is
$\sim1$ {\AA}/pixel. The fibre-coupling of each beam to two CCD 
chips introduced additional small gaps in the wavelength 
coverage in the regions 4360--4370~{\AA} and 7000--7020~{\AA}. 
The spectra were taken with 8~arcsec slits and the                  
spectral resolution was 5~{\AA} (FWHM).

Tresse et al. (1999) measured the 4000~{\AA} breaks and emission-line 
properties of 1671 galaxies with $15\leq \bj\leq17.15$ in the sample 
(the other 126 galaxies in the original sample were either brighter than 
$\bj=15$, or their spectra were blue-shifted, or they had published 
redshifts). They measured the fluxes of all lines with emission 
equivalent widths larger than 2~{\AA}, corresponding to a $3\sigma$ 
detection limit. Because of the wavelength gaps in the spectra, different
lines were measured for galaxies at different redshifts.  

We first exclude 79 galaxies with $\nii\lambda6583/\ha>0.63$, which are likely
to be AGN. The fluxes of \ha, \oii$\lambda{3727}$, \nii$\lambda6583$ and 
\sii$\lambda \lambda{6717,\, 6731}$ could be measured for 416 
galaxies. The fluxes of subsets of these lines are available for 588 
additional galaxies.  Of these, 289 galaxies have measurements of \ha, \oii,
and either \nii\ or \sii. Finally, 588 galaxies have no fluxes in emission
with equivalent widths larger than 2~{\AA}. We note that
since \hb\ and \oiii$\lambda{5007}$ fall in between the blue and red 
beams of the spectrograph for most objects, they are generally not 
observed. In this paper, we focus on the subsample of 705 non-Seyfert 
galaxies for which \ha, \oii, and at least one of \nii\ or \sii\ were
measured.

The absolute flux calibration of the original SAPM spectra is only 
accurate to $\sim20$ per cent because of the limited number of standard
stars observed each night (Singleton 2001). However, the {\em relative}
flux calibration, which is more critical for our analysis of emission
line ratios and equivalent widths is accurate to $\sim5$ per cent
(Tresse et al. 1999). The 8~arcsec-wide and 7~arcmin-long slit used in
the observations samples typically 45 per cent of the rest-frame projected
area of a galaxy at a \bj\ surface brightness level of 25 mag~arcsec$^{-2}$. 
To correct for this effect, Singleton (2001) computed the flux in the \bj\
band in each spectrum and derived an aperture correction for each galaxy
using the ratio of the spectral flux to the total \bj\ band flux of the 
galaxy. 

The above aperture 
correction should be accurate so long as the line emission and
the blue light are correlated over the entire galaxy. One worry is
that redshift surveys that use limited spectroscopic apertures may be
biased as a function of morphological type, because spectroscopic 
properties vary significantly from the inner (bulge-dominated) to
the outer (disc-dominated) regions of galaxies. Kochanek, Pahre \& 
Falco (2001) have shown, however, that the long-slit SAPM spectra 
sample the overall light distribution of a galaxy far better than the 
small-aperture (e.g. 2--3~arcsec fibre) spectra of other nearby
spectroscopic surveys.  They conclude that the spectral classification of
SAPM galaxies should not suffer from aperture bias. Also, Tresse et al.
(1999) showed that the average properties of galaxies in the SAPM sample are 
similar, in terms of line ratios (e.g. \niiha\ and \oiiha), to those in the 
integrated-spectrum sample of Kennicutt (1992b). As found by Kennicutt
(1992b), the main effect of disc undersampling is to reduce the strengths
of the various emission lines in roughly equal proportion.  Thus, the 
aperture corrections derived by Singleton (2001) should be appropriate. 
We expect, therefore, that the spectral analysis of the SAPM galaxy sample
presented below does not suffer from any significant aperture bias.

Observations from IRAS are also available for 279 galaxies in the sample
(Singleton 2001). Only 149 of these belong to our subsample of 705 
non-Seyfert galaxies with measurements of \ha, \oii, and one or both
of \nii\ and \sii.

In summary, we consider two samples in this paper: (1) a restricted sample
of  149 non-Seyfert galaxies with IRAS detections, for which fluxes and
equivalent widths of \ha, \oii, and one or both of \nii\ and \sii\ are 
available. We use this sample to compare star formation rates derived 
using different estimators based on the H$\alpha$, \oii\ and far-infrared
(FIR) luminosities of galaxies; (2) a larger sample of 705 non-Seyfert 
galaxies for which fluxes and equivalent widths of \ha, \oii, and one or 
both of \nii\ and \sii\ are available. We use this sample to study trends
in star formation rate and metallicity as a function of galaxy absolute 
\bj\ luminosity.

In this paper, we adopt a Hubble constant $H_0=70 $~km$\,$s$^{-1}$Mpc$^{-1}$.   

\section {Star Formation rates Derived using Standard Estimators}

In this section, we adopt the formulae given in K98 to convert H$\alpha$,
\oii, UV and FIR luminosities into star formation rates, and we check the
results for consistency.

\subsection{Standard SFR estimators}

The K98 calibration of the \ha\ luminosity emitted per unit SFR is derived
by using standard population synthesis models to calculate the ionizing 
radiation produced by young stars and by applying dust-free case~B 
recombination to this. For solar abundances and a Salpeter IMF (0.1--100 
M$_{\odot}$), K98 derives the following transformation between \ha\ 
luminosity and SFR:
\begin {equation}
\sfrha\,( \rm{M}_{\odot}\,\rm{yr}^{-1}) =
7.9\times 10^{-42} \lha\,(\rm{erg\,s}^{-1}).
\label{eqha}
\end {equation}

This calibration does not include the effects of attenuation by dust. 
Kennicutt (1983) and Niklas, Klein \& Wielebinski (1997) have used
observations of the integrated \ha\ and thermal radio fluxes of nearby 
spiral galaxies to derive a mean attenuation $\Aha=0.8-1.1$ mag for these
objects. Unless otherwise specified, we apply a `standard' attenuation
correction $\Aha=1$ mag to all galaxies in the sample. We emphasize that
this corrects for only part of the effects of dust on the \ha\ line, namely
the absorption by dust of \ha\ photons outside the \hii\ regions in which
they are produced. The correction neglects the diminution of the \ha\ line
caused by the absorption of ionizing photons by dust inside \hii\ regions
(Petrosian, Silk \& Field 1972; Mathis 1986; Charlot \& Fall 2000). More
refined attenuation corrections based on the departure of the observed
\hahb\ ratio of each galaxy from dust-free case~B recombination would 
suffer from the same limitation (we recall that \hb\ luminosities are not
available for SAPM galaxies). 

The luminosities of forbidden lines such as \oii\ are not directly coupled
to the ionization rate, and their excitation is known to be sensitive to
the abundance and the ionization state of the gas. However, it is claimed
that the excitation of \oii\ is sufficiently well behaved in observed galaxies
that it can be calibrated {\em empirically} through \ha\ as a quantitative
SFR tracer (Gallagher et al 1989; Kennicutt 1992b). K98 gives the formula
\begin {equation}
\sfroii\,( \rm{M}_{\odot}\,\rm{yr}^{-1}) =(1.4 \pm 0.4)\times 10^{-41} \loii\,
(\rm{erg\,s}^{-1}).
\label{eqoii}
\end {equation}
It is worth noting that, because this estimator is based on an empirical
calibration, the {\em relative} attenuation of the \ha\ and \oii\ lines is
automatically accounted for. We also apply a standard attenuation correction
${\rm \Aha}=1$ mag to the \oii-derived SFR estimates. 

The non-ionizing ultraviolet (UV) radiation from young stars is another
standard SFR estimator, especially in high-redshift galaxies where
optical emission lines are not always accessible. K98 gives the following 
transformation between UV luminosity \luvnu\ and SFR:
\begin {equation}
\sfruv\,(\rm{M}_{\odot}\,\rm{yr}^{-1}) = 
1.4\times10^{-28} \luvnu\,(\rm{erg\,s}^{-1} \rm{Hz}^{-1}).
\label{equv}
\end{equation}
This formula includes no dust correction and assumes that the
SFR has remained constant over timescales that are long compared
to the lifetimes of the dominant UV emitting population ($ < 10^8$ yr).
For the choice of a Salpeter (0.1--100 M$_{\odot}$) IMF, the UV luminosity
is practically independent of wavelength. In Section 3.2 below, we
address the uncertainty in the above relation that arises as a result of
the absorption of UV radiation by dust.

The FIR emission should provide an excellent measure of the SFR in dusty, 
circumnuclear starbursts. As discussed by K98, in the optically thick limit,
it is sufficient to model the bolometric luminosity of the stellar population,
the greatest uncertainty being the assumed age of the starburst.
K98 quote the following relation for continuous bursts of age 10--100 Myr:
\begin {equation}
\sfrfir\,(\rm{M}_{\odot}\,\rm{yr}^{-1}) =4.5 \times 10^{-44} 
\lfir\,({\rm erg\,s}^{-1})\,,
\label{eqfir}
\end {equation}
where $L_{\rm{FIR}}$ refers to the luminosity integrated over the full 
mid-infrared to submillimeter spectrum (8--1000 $\mu$m).

In more normal star-forming galaxies, the relation between SFR and FIR emission
is more complex. As discussed by K98, the contribution to dust heating from
older stars will lower the effective coefficient in Equation (\ref{eqfir}), 
whereas the lower optical depth of the dust will tend to increase the coefficient.
Buat \& Xu (1996) computed the total FIR luminosity emitted per 
unit SFR in spiral galaxies of type Sb to Irr using ultraviolet and 
IRAS observations of 152 nearby disc galaxies. For this sample, they derived a 
mean coefficient a factor $\sim 2$ larger than that given in Equation (\ref{eqfir}).
Furthermore, Misiriotis et al. (2001) have carried out a detailed study of the 
optical to FIR spectral energy distributions of nearby, edge-on galaxies of type
Sb-Sbc. Their data also indicate that naive application of Equation (\ref{eqfir})
to these galaxies will {\em underestimate} the SFRs by a factor of $\sim 2$. [Note
that despite these considerations, we will always adopt the expression in Equation
(\ref{eqfir}) above when computing \sfrfir.]

\subsection{Comparison of SFRs derived using different standard estimators}

The SFR estimators above are valid in opposite limits for the transfer
of line and continuum radiation through the ISM of galaxies. Equation~(\ref{eqha})
assumes that the galaxies are transparent. Standard attenuation corrections
are at best very approximate remedies. Equation~(\ref{eqoii}) simply rewrites 
Equation~(\ref{eqha}) using an empirically derived proportionality constant, and
so applies in the same limit (although with significant additional uncertainties).
Equation~(\ref{equv}), which also applies to transparent galaxies, traces the 
emission from stars that live longer than those dominating the ionizing radiation.
Equation~(\ref{eqfir}), on the other hand, is valid only for opaque galaxies.
It would thus be surprising if the four estimators agreed, even for galaxies 
with perfect measurements of \lha, \loii, \luvnu\ and \lfir.

To compare the star formation rates obtained using these different estimators,
we first estimate the total FIR luminosity $L_{\rm{FIR}}$ for the 149
galaxies in our restricted SAPM sample with IRAS flux densities
at 60 and 100 $\mu$m. We adopt the following standard procedure
(see for example Meurer et al. 1999). We first calculate the quantity $F_{\rm
FIR}=1.26 \times 10^{-11} [2.58f_{\nu}(60\,\mu{\rm m})+ f_{\nu} (100\,\mu{\rm
m})]\, \,$erg$\,$cm$^{-2}$s$^{-1}$ defined by Helou et al. (1988) and then 
compute the total far-infrared flux from $F_{\rm FIR}$ and $f_{\nu}(60 
\,\mu{\rm m}) /f_{\nu} (100\,\mu{\rm m})$ using the relation appropriate 
for dust with a single temperature and an emissivity proportional to 
frequency $\nu$. This prescription has been shown work well for starburst 
galaxies (see Fig.~2 of Meurer et al. 1999). It is expected to 
a good approximation for all galaxies in which the infrared emission peaks
at wavelengths between $60\,\mu$m and $100\,\mu$m (Helou et al. 1988).  
The uncertainties that arise from the conversion of IRAS fluxes to total
FIR flux are likely to be  modest in comparison to the uncertainties in the
conversion from $L_{\rm{FIR}}$ to star formation rate.

Fig.~1 shows the logarithmic difference between \sfrha, \sfroii\ and 
\sfrfir\ for the 149 SAPM galaxies in this sample as a function of 
absolute rest-frame \bj\ magnitude. In each panel, we indicate the 
median $\Delta\log(\sfr)$ offset for galaxies with $M(\bj)<-19$ 
(representing 80 per cent of the sample), along with the associated 
mean offset and the rms scatter (in parentheses). For both \ha\ and 
\oii, we have included the `recommended' mean attenuation correction 
$\Aha=1$ mag. As expected, the three estimators yield inconsistent results. 

Fig.~1a is consistent with the recent result by Jansen, Franx \& 
Fabricant (2001; see also Carter et al. 2001) that bright galaxies 
tend to have lower \oiiha\ ratios than less luminous galaxies. We find
that, for $M(\bj)<-19$, \oii-derived SFR estimates are typically 40 
per cent smaller than \ha-derived ones, while at fainter magnitudes,
the two estimators give similar results. At fixed magnitude, the rms 
scatter in \sfroii/\sfrha\ is a factor of $\sim 1.7$, although discrepancies
between \ha- and \oii-derived SFR estimates can reach a factor of 7. 
The scatter in this plot and the weak trend with magnitude can probably
be explained by variations in the effective gas parameters (ionization,
metallicity, dust content) of the galaxies. We note that $M(\bj)\sim
-19.5$ corresponds to the mean magnitude of the galaxies which K98 used
to calibrate \oii\ as a SFR estimator. 

The differences between the SFRs estimated using optical lines 
and those estimated using the total FIR luminosity are much more 
severe. Figs.~1b--1c show that \ha- and \oii-derived SFR estimates 
are typically a factor of $\sim3$ lower than FIR-derived ones,
{\em even though we have applied an attenuation of 1~mag at \ha\ to 
the data}. We note that, as discussed above, the K98 formula given
in Equation (\ref{eqfir}) may well underestimate the true star 
formation rate in normal star-forming galaxies. As we shall see in
Section 5, the discrepancy in Figs.~1b--1c is likely to arise from 
the conversion of optical emission-line luminosities into star 
formation rates.

Ultraviolet luminosities are not available for the SAPM galaxies. 
To compare \sfrha\  and \sfrfir\ to the star formation rates derived
using UV luminosities, we have analyzed 37 UV-selected, normal star-forming
galaxies from the sample of Bell \& Kennicutt (2001) and 18 UV-selected
starburst galaxies from the sample of Meurer et al. (1999). We selected
in the original samples all the galaxies with \ha, UV and IRAS $60\,\mu$m
and $100\,\mu$m fluxes. The published restframe UV fluxes correspond to
the mean flux density over a passband with a central wavelength of 
$\lambda_{\rm UV} \approx1600$~{\AA} and a width of a few hundred 
angstrom.  

Fig.~2 shows the logarithmic difference between \sfruv, \sfrha\ and
\sfrfir\ for the galaxies in these samples as a function of absolute 
rest-frame $B$ magnitude. For consistency, the \ha-derived SFRs in
Fig.~2a are not corrected for the attenuation of \ha\ photons by dust.
Starburst galaxies follow roughly the same trends as normal star-forming
galaxies in these diagrams, although they have somewhat lower UV-derived
SFRs at fixed $B$-band magnitude. This may be explained if UV photons are
more strongly attenuated by dust in starburst galaxies, or if long-lived
stars with ages $10^7$--$10^8\,$yr contribute a larger fraction of the
UV-luminosity in normal star-forming galaxies. Fig.~2b is also consistent 
with an increase of attenuation by dust with increasing luminosity in 
nearby starburst galaxies (e.g. Heckman et al. 1998). Overall, the 
similarity between the properties of starburst and normal star-forming
galaxies in Fig.~2 suggests that the starbursts do not strongly influence
the global ISM parameters of galaxies.         

The median difference between the UV- and \ha-derived SFRs, both uncorrected
for dust, is small, i.e. less than a factor of 2. This was already noted 
by Bell \& Kennicutt (2001) for their sample. This means that even though
the overall attenuation by dust can be large (as Fig.~2b demonstrates), it is
similar for \ha\ and ultraviolet photons. Ultraviolet photons are expected to be
more attenuated than \ha\ photons for a normal extinction curve. On the other
hand, the stars dominating the \ha\ emission are more obscured than those
dominating the UV emission, because they have lifetimes ($\la3\times10^6\,$yr)
shorter than the typical timescale for the dispersal of clouds in which
they form [the dispersal timescale of giant molecular clouds in
the Milky Way is estimated to be about $10^7\,$yr; Blitz \& Shu (1980)]. 
Similar results are found for nearby starburst galaxies,  where the attenuation
inferred from the \hahb\ ratio is typically {\em higher} than that inferred 
from the ultraviolet and optical spectral continuum (e.g., Fanelli, O'Connell
\& Thuan 1988; Calzetti, Kinney \& Storchi-Bergmann 1994; Calzetti 1997; Poggianti
et al. 1999; see Charlot \& Fall 2000). In summary, our results show that even
when the attenuation is strong,  the  ratio of the emergent ultraviolet to 
\ha\ luminosities can be close to that in the dust-free case, leading to small
{\em apparent} discrepancies between the uncorrected UV- and \ha-derived SFRs.  

\section{The Charlot \& Longhetti Model}

In the previous section, we investigated a number of standard SFR estimators
based on the \ha, \oii, UV and FIR luminosities of galaxies. The different 
estimators give inconsistent results. These results highlight the need for a
physically realistic model to interpret the combined signatures of stars, gas,
and dust in the integrated spectra of galaxies. Such a model was recently 
developed by Charlot \& Longhetti (2001; hereafter CL01).

\subsection{Description of the model}

The CL01 model is based on a combination of the latest version of the Bruzual \& 
Charlot (1993) population synthesis code and the Ferland (1996, version C90.04) 
photoionization code. Stars are assumed to form in interstellar `birth clouds' 
(i.e. giant molecular clouds).  After $10^7\,$yr, young stars are assumed to 
disrupt their birth clouds and migrate into the `ambient
ISM'. The line emission from a galaxy arises both from the \hii\ regions surrounding
young stars and from the diffuse gas ionized by photons that have leaked into the 
ambient ISM. In the CL01 model, these different gas components are combined and are
described in terms of `effective' gas parameters. This is motivated by the fact that the 
optical-line ratios in the integrated spectra of nearby spiral galaxies are similar
to those in the spectra of individual \hii\ regions (Kobulnicky et al. 1999; see CL01
for more detail). The spectral evolution of a galaxy is obtained by convolving the
spectral evolution of a single stellar generation with the SFR function. Only stars 
younger than about $3\times10^6\,$yr contribute significantly to the line        
emission. 

The main adjustable parameters that affect the emission from the stars in a
galaxy are the IMF, the SFR and the stellar metallicity $Z$. The main parameters
affecting the emission from the photoionized gas are the effective interstellar
metallicity \zav, the zero-age effective ionization parameter \uavo, the 
effective dust-to-heavy element ratio \xav\ and the effective gas density \nav.
\zav, which was originally called `effective gas metallicity' in CL01, includes
all the heavy elements in the gas and solid (i.e. dust) phases. It is taken to 
be the same as the metallicity of the ionizing stars, i.e. $\zav=Z$. \uavo\ 
describes the {\em typical} ionization state of photoionized gas in the galaxy.
\xav\ is defined as the mass fraction of heavy elements locked into dust grains
in the ionized gas. The other main parameter of the model is the age of the galaxy $t$. 
In all models, we adopt a Salpeter IMF truncated at 0.1 and 100~$M_\odot$ and we
fix ${\rm \nav} \approx 30\,$cm$^{-3}$, a value that is typical of the gas density
in Galactic and extragalactic \hii\ regions.  Changes in \nav\ have a 
minor effect on the conversion between line luminosities and SFR (see CL01).
CL01 calibrated the emission-line properties of their model using the observed
\oiiihb, \oiioiii, \siiha\ and \niisii\ ratios of a representative sample of
92 nearby spiral and irregular, starburst and \hii\ galaxies.

The CL01 model also includes the  prescription for the absorption of photons 
from \hii\ regions and from older stars that was introduced  by Charlot \& Fall
(2000).  In this prescription,  line and continuum photons are attenuated
differently because of the finite lifetimes of the stellar birth clouds.
The effective absorption curve is given by the following
formula\footnote{In Charlot \& Fall (2000), $\hat{\tau}_V$ is used to denote
the {\em total} effective $V$-band optical depth, i.e. the sum of the effective 
optical depths in the \hii\ regions, the surrounding \hi\ regions of the
birth clouds and the ambient ISM ($\htauv=\htauvhii+\htauvhi+\htauvism$).
In the CL01 model, the absorption in the \hii\ regions is computed by the 
photoionization code, and the Charlot \& Fall (2000) model is used with 
$\hat{\tau}_V^{\rm HII}=0$ to compute the subsequent absorption in the 
surrounding \hi\ regions of the birth clouds and the ambient ISM. Thus, 
in CL01 and here, $\hat{\tau}_V$ represents the effective $V$-band optical
depth in the `neutral' ISM, i.e. $\htauv=\htauvhi+\htauvism$.}
\begin{eqnarray}
\hat{\tau}_\lambda(t')=\cases{
\hskip0.20cm\hat{\tau}_V\left(\lambda/{5500\,{\rm \AA}}\right)^{-0.7}\,,
&for $t'\leq 10^7$ yr,\cr
{{\mu\hat{\tau}_V}}\left(\lambda/{5500\,{\rm \AA}}\right)^{-0.7}\,,
&for $t'>10^7$ yr\,,\cr}
\label{taueff}
\end{eqnarray}
where $t'$ is the age of any single stellar generation. The wavelength 
dependence of the effective absorption curve $\hat\tau_\lambda$ is 
constrained by the observed relation between the ratio of FIR to UV 
luminosities and the UV spectral slope for starburst galaxies. The
age $10^7\,$yr corresponds to the typical lifetime of a stellar 
birth cloud (i.e. giant molecular cloud). The adjustable parameter $\mu$
defines the fraction of the total dust absorption optical depth of the
galaxy contributed by the ambient ISM. Charlot \& Fall (2000) showed that
$\mu \approx1/3$ is needed to reproduce the observed mean relation 
between \hahb\ ratio and ultraviolet spectral slope (defined near 1600~{\AA})
for starburst galaxies. The scatter in the relation is reproduced if 
$\mu$ ranges from values of $\sim 1$ (i.e., all the radiation from the
\hii\ regions leaks into the ambient ISM) to $\sim 1/5$ in different 
galaxies. In this prescription, the attenuation of the \ha\ line 
radiation, which is produced by stars younger than $\sim3\times10^6\,$yr,
can be expressed simply as $\Aha=1.086\,\htauv\,(\lamha/ {5500\,{\rm 
\AA}})^{-0.7}$ ($\lamha={6563 \,{\rm \AA}}$). The attenuation \Av\ of
the 5500~{\AA} continuum radiation, which is produced by stars both younger
and older than $10^7\,$yr, depends on the star formation history.

The different parameters of the CL01 model each have a specific influence
on the various line and continuum spectral features of galaxies. This allows
one to derive `physical' quantities such as star formation rates,  
gas-phase oxygen abundances and  effective dust absorption optical depths from the
optical spectra. The constraints will be more accurate if more spectral features
are available. 

\subsection{Application to the SAPM survey}

For the SAPM galaxies, the spectral features available to constrain physical
quantities are the fluxes and equivalent widths of \ha, \oii, and one or both
of \nii\ and \sii. For a subset of the galaxies, the total FIR luminosity is 
also available. The parameters we wish to constrain are the following. For the 
gas, they are the interstellar metallicity, the ionization parameter, the 
dust-to-heavy element ratio, the effective dust absorption optical depth of 
the `neutral' ISM (i.e. the birth clouds plus the ambient ISM), and the fraction
of this contributed by the ambient ISM.  For the stars, they are the age of the
galaxy and the star formation history (see Table 1). 

For the SAPM sample, the primary constraint on the interstellar metallicity 
comes from the \niisii\ ratio, if both metal lines are available. If only
one line is available, the metallicity will be set by \niiha\ or by \siiha.
The \siiha\ ratio also constrains the dust-to-heavy element ratio \xav, and
the \oiisii\ ratio constraints the ionization parameter \uavo. The 
effective absorption optical depth of the dust \htauv\ is constrained 
primarily by the \oiiha\ ratio. As can be seen, there are at most 3 independent
line ratios available to constrain 4 gas parameters, so the resulting errors 
on some of these parameters will be large. The constraints on the effective 
dust absorption optical depth, for example, would be far tighter if the 
\hahb\ ratio were available. In fact, for the subset of galaxies with 
far-infrared luminosities, the \lha/\lfir\ ratio greatly improves the
constraints on \htauv. The parameters $\mu$ and $t$ and the history of star
formation SFR(t) are constrained only weakly by the measured equivalent 
widths of the various lines, the \bj\ absolute magnitude and the 4000~{\AA} break.
This is why we do not show any results for the parameters pertaining to 
long-lived stars in this paper. The specific influence of the various parameters
on observables allows one to construct `optimal' estimators of the star
formation rate, the gas-phase oxygen abundance and the dust absorption
optical depth based on observed spectral features.
This was the approach adopted by CL01 (their table 2).\footnote{We note that
the CL01 model reproduces equation~(\ref{eqha}) to within a few percent
if $\xav=\htauv=0$ (i.e. in the case of a completely transparent galaxy),
and if the effects of stellar \ha\ absorption are ignored. 
Equation~(\ref{eqfir}) is recovered if $\mu\htauv\gg1$.} More accurate 
constraints and better estimates of the errors in the derived quantities may
be obtained if the models are used to fit the spectral properties of each 
individual galaxy. This is the procedure we adopt in this paper.

Following CL01, we first select a set of models that span   
the full range of observed optical-line ratios, ultraviolet and 
far-infrared luminosities of nearby star-forming galaxies. The parameters
of these models are listed in Table~\ref{partable}. They are the same as
those listed in table~1 of CL01, except that we have now extended  
the ranges of metallicities and of effective dust absorption optical depths.  
We construct a grid of $1.6\times10^8$ models by systematically varying the
parameters listed in Table~\ref{partable}. In practice, this is done by 
interpolating from the coarse grid of $7.1\times 10^5$ models computed by CL01
(with extensions in \zav\ and \htauv). For each galaxy in our sample, we 
evaluate the \chisq\ goodness of fit of all the models in the grid.
The SFR, \oh\ and \htauv\ of the mininum-\chisq\ model define
our best estimates of the star formation rate, the gas-phase oxygen 
abundance and the effective optical depth of the dust in the galaxy. The
maximum ranges in SFR, \oh\ and \htauv\ among the models within 
$\Delta\chisq=1$ of the minimum provide estimates of the uncertainties 
in these quantities.

\begin{table}
 \caption{Parameters of the models reproducing the observed integrated
spectral properties of nearby star-forming galaxies. Two timescales 
are adopted for the exponentially declining star formation rate 
SFR$(t)$: 0.1 and 6.0~Gyr. All models have $\tilde{n}_{\rm H}=
30~$cm$^{-3}$ and a Salpeter IMF truncated at 0.1 and 100~$M_\odot$.}
 \label{partable}
 \begin{tabular}{@{}ll}
Parameter & Range \\
${\rm \uavo}$ & $10^{-3.0}$--$10^{-1.5}$ \\
${\rm \zav}/Z_\odot$ & 0.2--4.0 \\
${\rm \xav}$ & 0.1--0.5 \\
$\hat{\tau}_V$ & 0.01--4.0 \\
$\mu$ & 0.2--1.0 \\
SFR$(t)$ & constant, exponentially declining \\
$t$/yr & $10^7$--$10^{10}$ \\
\end{tabular}
\end{table}

\section{Results}

We now present the star formation rates, gas-phase oxygen abundances
and dust absorption optical depths we derive for the SAPM sample
using the CL01 model described in Section~4. 

In Fig.~3, we present results for the 149 SAPM galaxies with fluxes and
equivalent widths of \ha, \oii, and one or both of \nii\ and \sii, as 
well as IRAS 60 and 100 $\mu$m fluxes. In this plot,
we do not use the IRAS fluxes to constrain the SFRs. We compare the SFRs 
derived using all available optical lines with those derived from the
\ha, \oii\ or FIR luminosities alone using the K98 formulae given in
Equations (\ref{eqha}), (\ref{eqoii}) and (\ref{eqfir}). The filled 
squares represent galaxies with measurements of all four optical lines, 
while the open circles and the open squares are galaxies for which \ha,
\oii, and either \nii\ or \sii\ were measured. The error bars indicate the
uncertainties in the SFRs derived using the CL01 models, as discussed 
in Section~4.2. Once again we have applied a standard attenuation ${\rm
\Aha}=1$ mag to the SFRs estimated using the K98 formulae for \ha\ and
\oii. 

We find that the SFRs derived using \ha\ are lower than those 
obtained using the CL01 model. The median discrepancy tends to
increase with \bj\ absolute luminosity and amounts to a factor
of $\sim3$ at $M(\bj)<-19$ (Fig.~3a). The discrepancy is similar,
but the scatter is larger for the SFRs derived using \oii\ (Fig.~3b). 
Interestingly, the CL01 SFRs are in much better agreement with 
those derived from the FIR luminosities, {\em even though we did not
use these measurements to constrain the models} (Fig.~3c).

Fig.~4 shows what happens if we add the requirement that models 
reproduce the observed FIR luminosities of the galaxies. The
uncertainties are much smaller in this case because fewer models 
fit the data. The median error on the SFR decreases from a factor
of $\sim 1.8$ in Fig.~3 to a factor of $\sim 1.3$ in Fig.~4. The 
SFRs derived using the CL01 model now agree surprisingly well with 
those derived using the K98 FIR-luminosity estimator, with a 
scatter of only 50 per cent. 

Why are the CL01 SFR estimates so different from the SFRs derived using 
the K98 \ha\ formula plus a standard attenuation correction of $\Aha=1$ mag?   
Why are they so similar to those derived using the K98 FIR-luminosity
formula for optically thick starburst galaxies?  The distribution of \ha\ 
attenuations that we derive for the SAPM galaxies is similar to that 
found in other samples of nearby star-forming galaxies. To illustrate
this, Fig.~5a compares the distribution of \Aha\ with those derived by 
Sullivan et al. (2000) and Bell \& Kennicutt (2001) using \ha, \hb\ and
thermal radio continuum observations of nearby, UV-selected star-forming
galaxies [we have converted \Av\ from figure 1 of Sullivan et al. (2000)
into $\Aha=0.78\Av$, consistent with the recipe these authors used in the
first place to compute \Av\ based on the \hahb\ ratio]. The median \ha\ 
attenuation in our sample, $\Ahamed \approx1.1$, is only slightly larger
than those of the Sullivan et al. (0.8) and Bell \& Kennicutt (0.9) 
UV-selected samples.

It is important to distinguish here between \ha\ and $V$-band attenuation.
Although most galaxies in our sample are optically thick at \ha, they
are generally not in the $V$ band. This is because the $V$-band continuum
emission is dominated by stars that live longer, and hence, that are less
obscured, than those dominating the \ha\ line emission (Section 4.1). Fig.~5b
shows the distribution of $V$-band attenuations for our sample. This has
a median $\Avmed\approx0.8$. The results of Fig.~4c then suggest that dust 
heating by long-lived stars roughly compensates the relatively low dust
optical depths of the galaxies, in such a way that the ratio of FIR 
luminosity to SFR is similar to that predicted by the K98 formula for 
optically thick starburst galaxies.

We now show that differences between the CL01 and the K98 \ha\ estimates of
the SFR arise as a result not of one, but of several simplifying approximations
in the K98 formula. It is useful to step back and compare the CL01 SFRs 
with those derived from Equation (\ref{eqha}), {\em before} applying
any correction for \ha\ attenuation. This is shown in Fig.~6a. Rather 
than applying a standard attenuation correction of 1 mag, we now correct
the \ha-derived SFRs of individual galaxies using our estimated attenuations
shown in Fig.~5 above. As expected, this reduces the scatter in the diagram,
but an offset of a factor of $\sim2$ remains between the CL01 and K98 estimates
(Fig.~6b). A similar discrepancy would presumably be found if attenuation were
derived, for example, using the \hahb\ ratio.

An effect that is included in the CL01 model, but not in the K98 formula,
is the absorption of ionizing photons by dust inside \hii\ regions (Section
4.1 above). We find that, for the 149 galaxies in Fig.~6, the mean and rms
scatter of the fraction of ionizing photons absorbed by dust correspond to
$\fabs\approx22\pm20$ per cent. This is constrained primarily by the 
\siiha\ and \oiisii\ ratios of the galaxies, which control the dust-to-heavy
element ratio and ionization parameter of the gas (Section 4.2 and CL01). 
For reference, Degioia-Eastwood (1992) finds $\fabs\approx35\pm11$ per 
cent based on optical, far-infrared and radio observations of 6 \hii\ 
regions in the Large Magellanic Cloud. In Fig.~6c, we show the effect of
correcting the \ha-derived SFRs of individual galaxies in Fig.~6b for the
fraction of ionizing photons lost to dust. Both the scatter and the offset
relative to the CL01 estimates of the SFR are reduced, but there remains
an offset of a factor of $\sim1.5$.

Another effect that is included in the CL01 model, but ignored by the
K98 formula, is the contamination of \ha\ emission by stellar absorption.
For the galaxies in Fig.~6, the mean and rms scatter of the equivalent 
width of H-Balmer absorption correspond to $\abswha\approx\abswhb\approx
-6.5\pm1.3$ {\AA}. This compares well with the rough estimate by Kennicutt 
(1992b) of $\abswhb\sim-5$~{\AA} for a sample of 90 nearby spiral 
galaxies. The correction for stellar absorption is therefore quite 
important for galaxies with small observed \ha\ emission equivalent 
widths. This is shown in Fig.~6d, where we correct the \ha-derived SFRs
of individual galaxies in Fig.~6c for stellar absorption. This additional
correction brings the \ha-derived SFRs into agreement with the CL01 
model. Some minor scatter remains ($\sim15$ per cent), mainly because 
the CL01 model also accounts for the decrease in the ionization rate for
increasing stellar metallicity (stars are assumed to have the same
metallicity as the gas; see Section 4.1 above). The K98 prescription, on
the other hand, assumes fixed solar metallicity. We conclude from Fig.~6,
therefore, that the neglect of several important physical effects in the
derivation of the K98 \ha\ SFR estimator leads to significant 
underestimates of the true rate of star formation occurring in typical
galaxies.

Fig.~7 presents the estimated SFRs and gas-phase oxygen abundances of
the 705 SAPM galaxies with fluxes and equivalent widths of \ha,
\oii, and one or both of \nii\ and \sii. The SFR in Fig.~7a is plotted 
per unit \bj\ luminosity. The median error in SFR/$L(\bj)$ is about 80
per cent, consistent with the results of Fig.~3. Unlike the \ha\ and \oii\
equivalent widths (Fig.~8), the quantity SFR/$L(\bj)$ exhibits no 
dependence on galaxy absolute \bj\ luminosity. Many previous studies have
noted that galaxies with strong \ha\ or \oii\ equivalent widths dominate
the faint end of the galaxy luminosity function (e.g., Zucca et al 1997; 
Loveday, Tresse \& Maddox 1999; Bromley et al 1998; Blanton \& Lin 2000;
Christlein 2000; Balogh et al. 2001). Our results show that this effect 
must arise in part because the emission lines in low-luminosity galaxies are
less attenuated by dust, rather than because faint emission-line galaxies form
stars at higher relative rates than their more luminous counterparts. The 
gas-phase oxygen abundances of the galaxies in our sample increase from 
$12+\log(\oh)\approx 8.5\pm0.3$ at $M(\bj) \sim-18$ to $12+\log(\oh)\approx
8.8\pm0.2$ (i.e. solar abundance) at $M(\bj)\sim-21$, indicating that more
luminous galaxies are more metal-rich, although with a large scatter. This
is agreement with other studies of trends in galaxy metallicity from 
observations of \hii\ regions in nearby star-forming systems (e.g. Zaritsky,
Kennicutt \& Huchra 1994). 

\section{Summary and Discussion}

We have compared a number of different methods for estimating the star formation
rates of galaxies using a variety of spectral indicators. Our primary sample is 
drawn from the Stromlo-APM redshift survey and consists of 149 galaxies with 
fluxes and equivalent widths of \ha, \oii, and one or both of \nii\ and \sii,
as well as IRAS 60 and 100 $\mu$m fluxes.

We first compared the star formation rates obtained using standard estimators
based on the \ha, \oii\ and FIR luminosities of the galaxies. These estimators
were derived in inconsistent limits using incomplete models of the reprocessing
of radiation from young stars, but they have, nevertheless, been used by many
authors to study star formation as a function of galaxy luminosity, morphological
type, environment and redshift. Not surprisingly, the three estimators give 
discrepant results. The star formation rates estimated using the far-IR fluxes
are a factor of $\sim 3$ higher than those estimated using \ha\ or \oii, even 
after the usual mean attenuation correction of $\Aha=1$ mag is applied to the data.

We then derived star formation rates for the same galaxies using estimators 
constructed from new models developed by CL01, which include a physically
consistent treatment of the effects of stars, gas and dust. The CL01 star 
formation rates agree much better with those derived from the FIR fluxes
than with those derived from the \ha\ or \oii\ fluxes. We showed that the 
discrepancy between the CL01 and \ha\ estimates of star formation arises primarily
because the standard \ha\ estimator does not include the absorption of ionizing 
photons by dust in \hii\ regions and the contamination of \ha\ emission by stellar
absorption. The strengths of these two effects as estimated by the models are in
good agreement with independent observations of normal, nearby star-forming galaxies.
The agreement with the FIR estimates is at least in part a fluke; in this case
the two major deviations from the assumed opaque-starburst limit have opposite
sign and approximately cancel for our SAPM sample.

Our results have important implications for estimates of the total integrated 
star formation density and its evolution with redshift (Madau et al 1996). Most
estimates of star formation rate densities at low redshift are based on \ha\ 
measurements (Gallego et al. 1995; Tresse \& Maddox 1998). If these studies
underestimate the total amount of star formation in the local Universe by 
factors of $\sim 3$, the resulting increase of the SFR density to high 
redshifts, which is generally inferred using different estimators, could be 
considerably less than commonly believed (see also Cowie, Songaila \& Barger
1999). We plan to investigate this in more detail in a future paper.

Finally, we have applied the CL01 models to 705 SAPM galaxies with fluxes 
and equivalent widths of \ha, \oii, and one or both of \nii\ and \sii\
to study trends in star formation rate and in metallicity as a 
function of galaxy absolute \bj\ magnitude. For this sample, we find that the
star formation rate per unit luminosity is independent of absolute magnitude.
The gas-phase oxygen abundance does increase with galaxy absolute \bj\ 
luminosity, although the scatter in metallicity at fixed magnitude is large.
We find that luminous galaxies have abundances close to solar on average. 

We note that the star formation rate and metallicity estimates derived using the CL01 
models are based on emission-line ratios. In order to obtain  reliable estimates
of these quantities, it is critical that the spectra be accurately flux-calibrated.
The large wavelength coverage and the expected spectro-photometric accuracy
of the galaxy spectra in the Sloan Digital Sky Survey (York et al 2000)
and the Two Degree Field survey (Folkes et al 1999) are ideally suited to this
kind of analysis. These surveys include measurements of key emission lines
not available for the SAPM (e.g. \oiii\ and \hb), which will greatly strengthen
the constraints on the effective ionization parameter and the 
effective dust absorption optical depth, even for galaxies that do not have
measured far-infrared fluxes. In addition, it will be possible to {\em test}
the CL01 models in detail using the additional spectral information.

In summary, both the models and the data are now available that will allow detailed
quantitative studies of the physical properties of galaxies as a function of 
luminosity, type and environment. This can only lead to an improved understanding 
of the processes that were important in the formation of the galaxies we see today.

\section*{Acknowledgments}

We thank R. Genzel and R. Kennicutt for valuable discussions that helped
shape this work. We also thank C. Singleton for making his data available
before publication.

\begin{figure*}
\epsfxsize=15cm
\hfill{\epsfbox{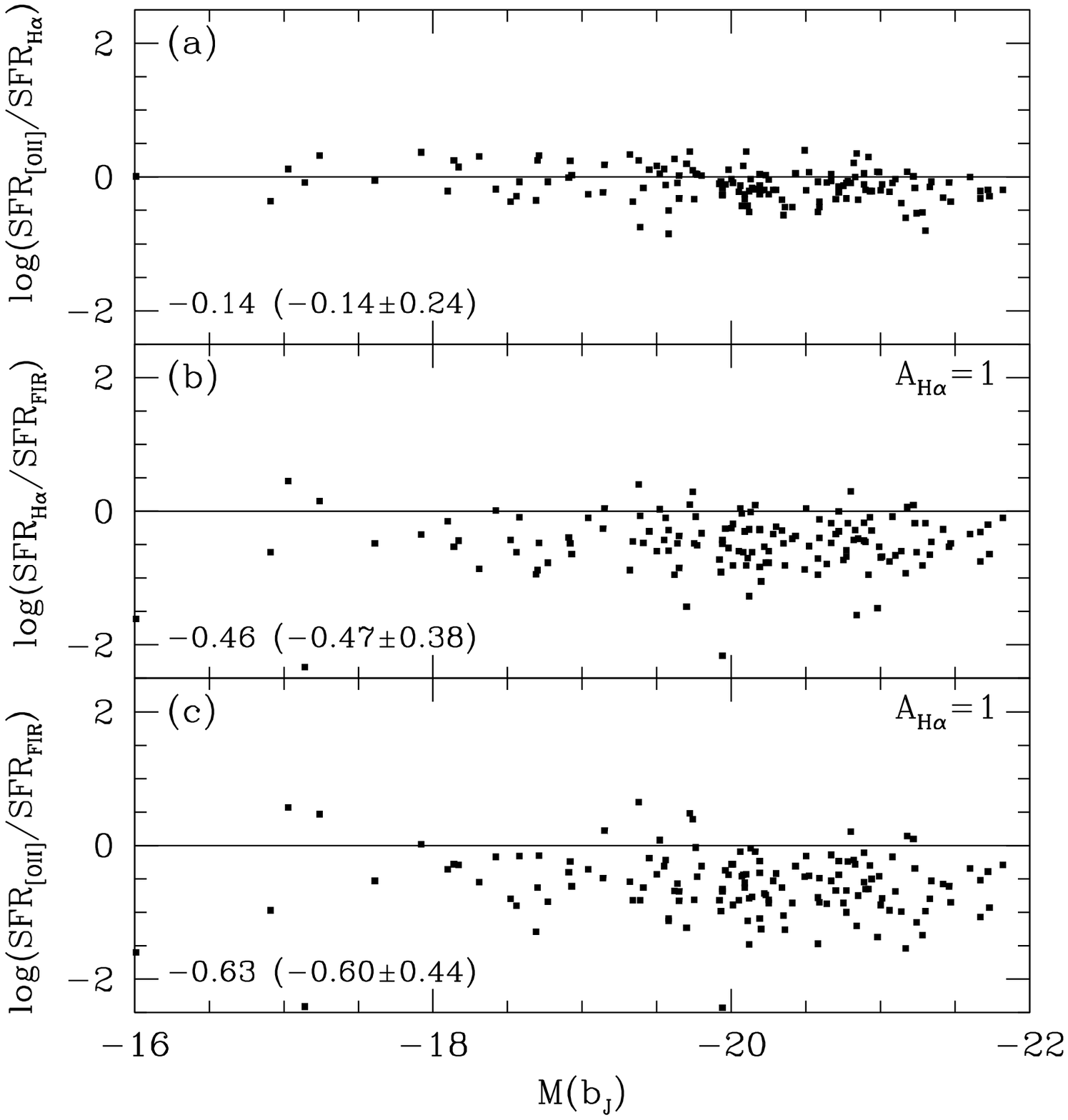}}
\caption{
Logarithmic difference between \sfrha, \sfroii\ and \sfrfir\ [equations 
(\ref{eqha}), (\ref{eqoii}) and (\ref{eqfir})] plotted against absolute rest-frame
\bj\ magnitude 
for the 149 SAPM galaxies with fluxes and equivalent widths of \ha, \oii, and
one or both of \nii\ and \sii, as well as IRAS 60 and 100$\,\mu$m fluxes. We have 
applied a standard attenuation correction of $\Aha=1$ mag to \sfrha\ and \sfroii.
In each panel, the median $\Delta\log(\sfr)$ offset for galaxies with $M(\bj)<-19$
is indicated along with the associated mean offset and the rms scatter 
(in parentheses).}
\end{figure*}

\begin{figure*}
\epsfxsize=15cm
\hfill{\epsfbox{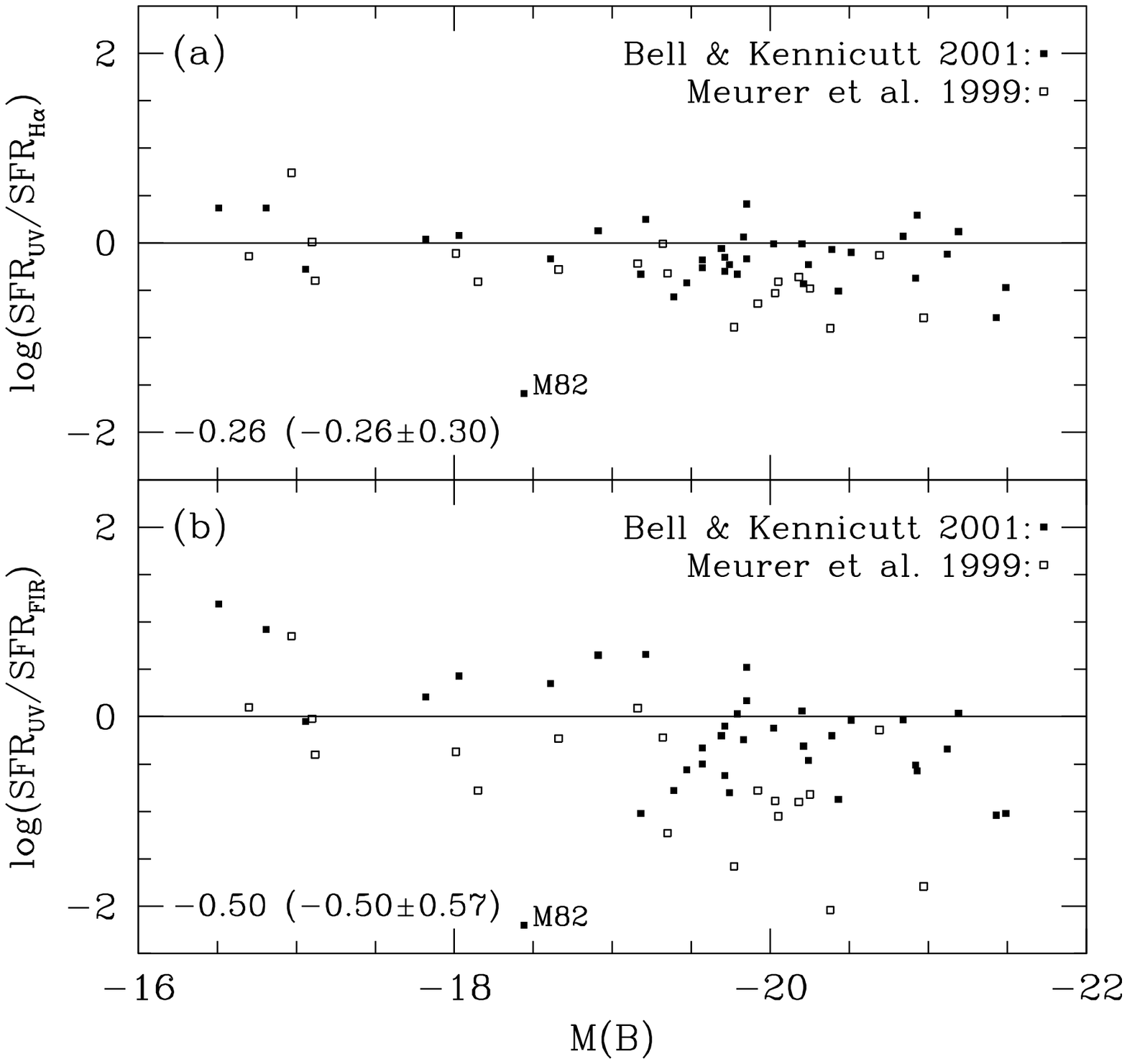}}
\caption{
Logarithmic difference between \sfruv, \sfrha\ and \sfrfir\ [equations 
(\ref{eqha}), (\ref{equv}) and (\ref{eqfir})] plotted against absolute rest-frame
$B$ magnitude for the galaxies with \ha, UV (1600~{\AA}) and IRAS 60 and 100 $\mu$m
flux measurements in the samples of Bell \& Kennicutt (2001; solid squares) and Meurer 
et al. (1999; open squares). In each panel, the median $\Delta\log(\sfr)$ offset for
galaxies with $M(\bj)<-19$ is indicated  along with the associated mean offset and 
the rms scatter (in parentheses).}
\end{figure*}

\begin{figure*}
\epsfxsize=15cm
\hfill{\epsfbox{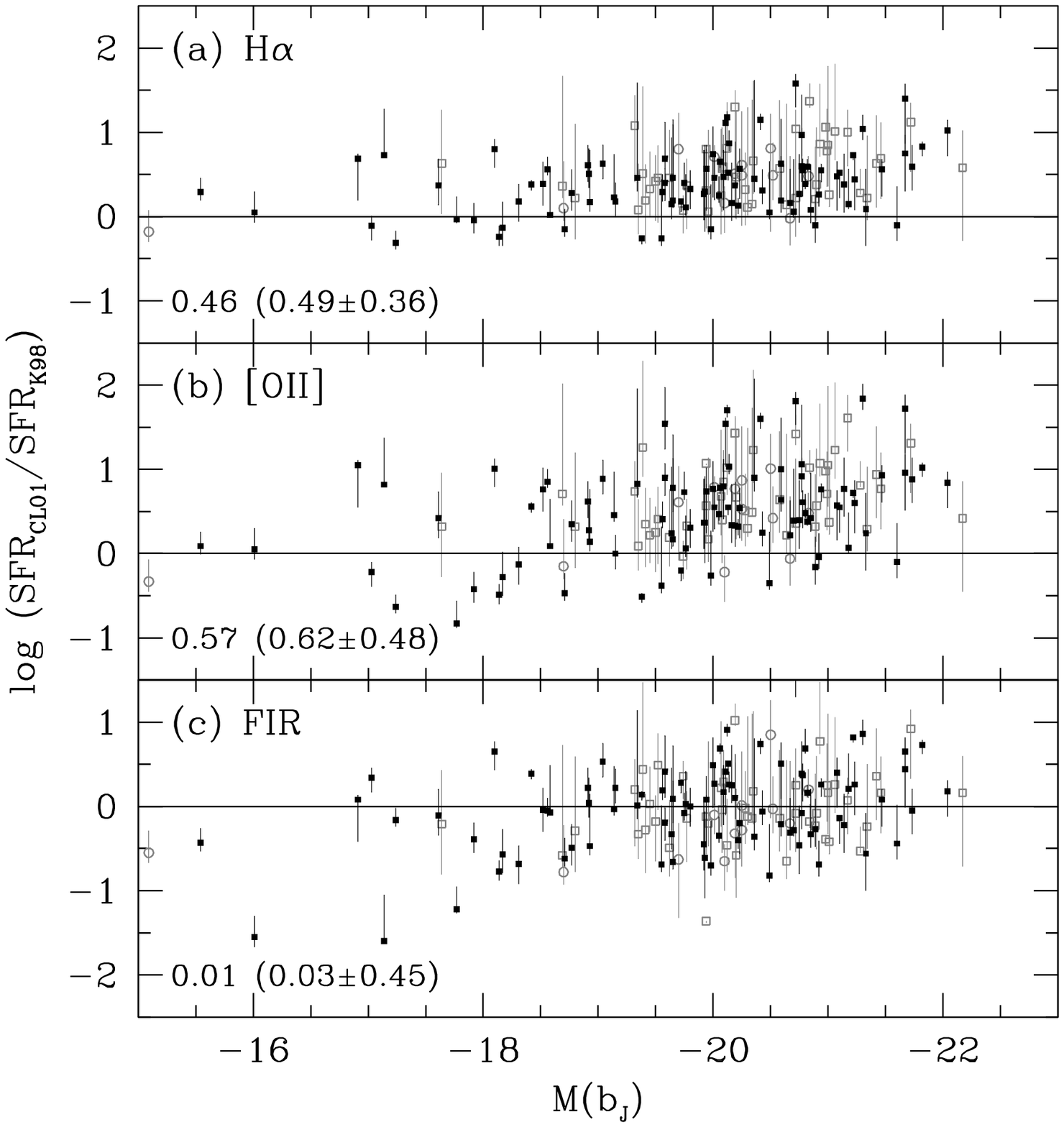}}
\caption{
Logarithmic difference between the CL01 SFR estimates and (a) \sfrha, (b) \sfroii\ 
and (c) \sfrfir\ [equations (\ref{eqha}), (\ref{eqoii}) and (\ref{eqfir})]
plotted against absolute
rest-frame \bj\ magnitude for the 149 SAPM galaxies with fluxes and equivalent 
widths of \ha, \oii, and one or both of \nii\ and \sii, as well as IRAS 60 and 
100$\,\mu$m fluxes. The CL01 models are required to fit the observed luminosities and
equivalent widths of the optical emission lines and the absolute \bj\ magnitude. The
filled squares represent galaxies with measurements of all four optical lines, the open
circles galaxies with \ha, \oii\ and \nii\ measurements, and the open squares galaxies
with \ha, \oii\ and \sii\ measurements. The error bars indicate the uncertainties in
the SFRs derived using the CL01 models, as discussed in Section~4.2. We have applied a 
standard attenuation correction of $\Aha=1$ mag to \sfrha\ and \sfroii. In each panel,
the median $\Delta\log(\sfr)$ offset for galaxies with $M(\bj)<-19$ is indicated along
with the associated mean offset and the rms scatter (in parentheses).} 
\end{figure*}

\begin{figure*}
\epsfxsize=15cm
\hfill{\epsfbox{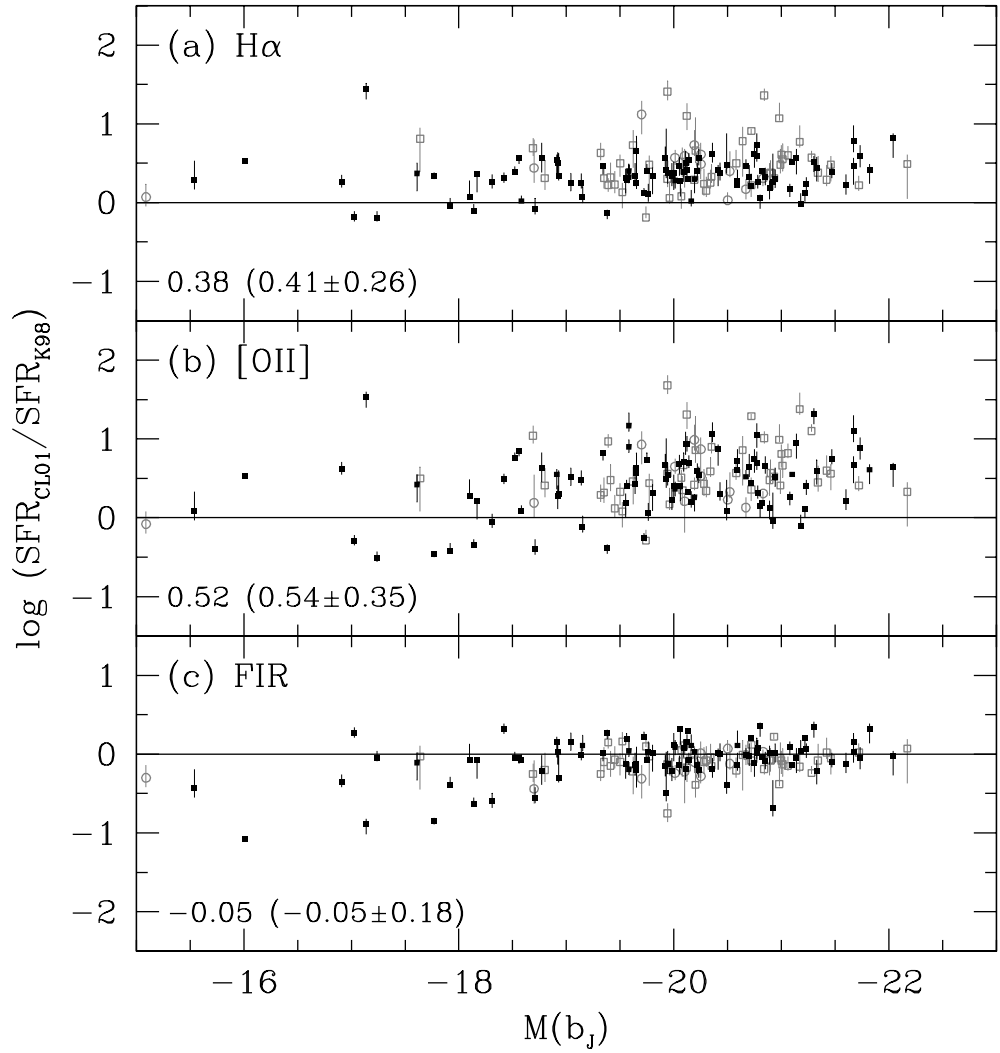}}
\caption{Same as Fig.~3, except that the CL01 models are now required to fit the 
total FIR luminosity (as defined in Section~3), in addition to the observed luminosities
and equivalent widths of the optical emission lines and the absolute \bj\ magnitude.}
\end{figure*}

\begin{figure*}
\epsfxsize=15cm
\hfill{\epsfbox{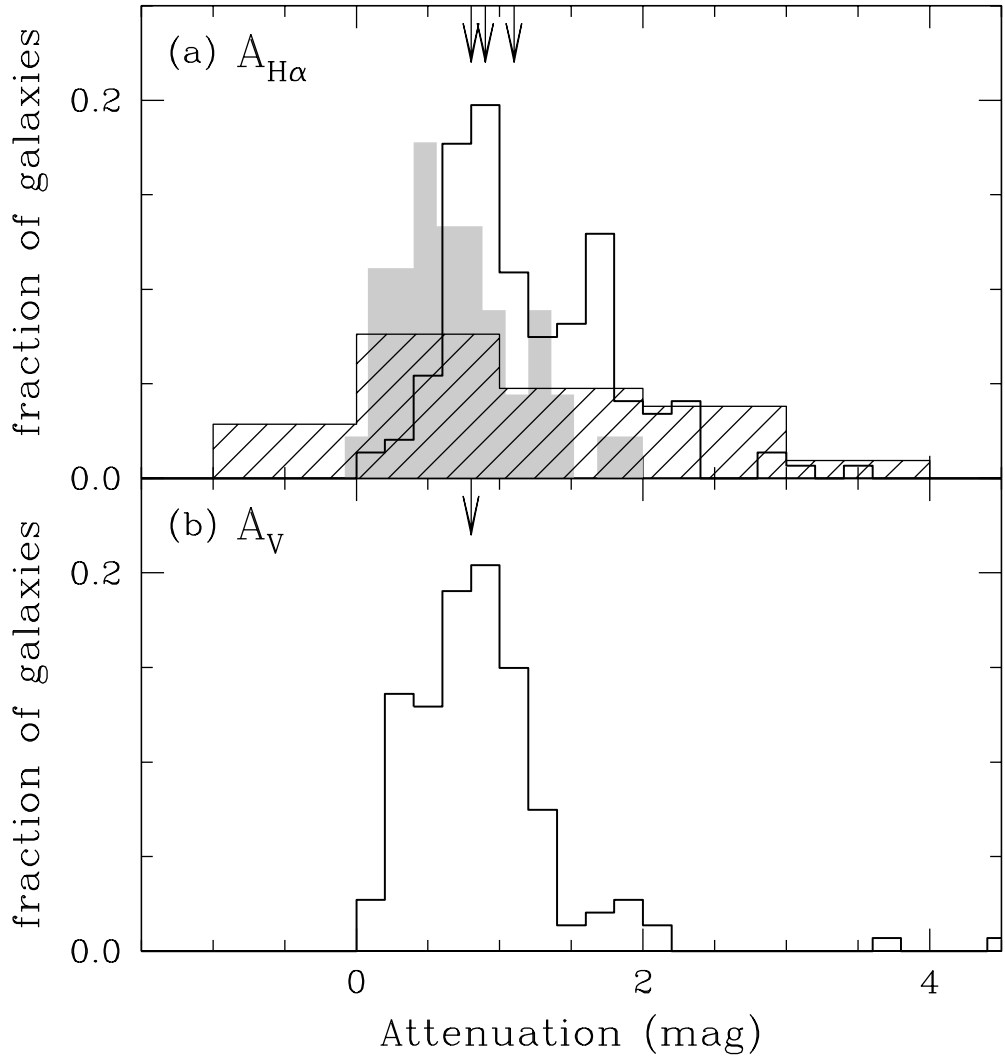}}
\caption{
(a) The distribution of \ha\ attenuations derived for the 149 SAPM galaxies with
fluxes and equivalent widths of \ha, \oii, and one or both of \nii\ and \sii, 
as well as IRAS 60 and 100$\,\mu$m fluxes (plain histogram). For comparison, 
results from Sullivan et al (2000) and from Bell \& Kennicutt (2001) are shown 
as shaded and hatched histograms, respectively. Arrows indicate the median 
attenuations in the Sullivan et al. (2000, $\Ahamed=0.8$), the Bell \& Kennicutt 
(2001; $\Ahamed=0.9$) and the SAPM ($\Ahamed=1.1$) samples. (b) The distribution
of $V$-band attenuations derived for the same sample of 149 SAPM galaxies as in
(a). An arrow indicates the median attenuation $\Avmed=0.8$.}
\end{figure*}

\begin{figure*}
\epsfxsize=15cm
\hfill{\epsfbox{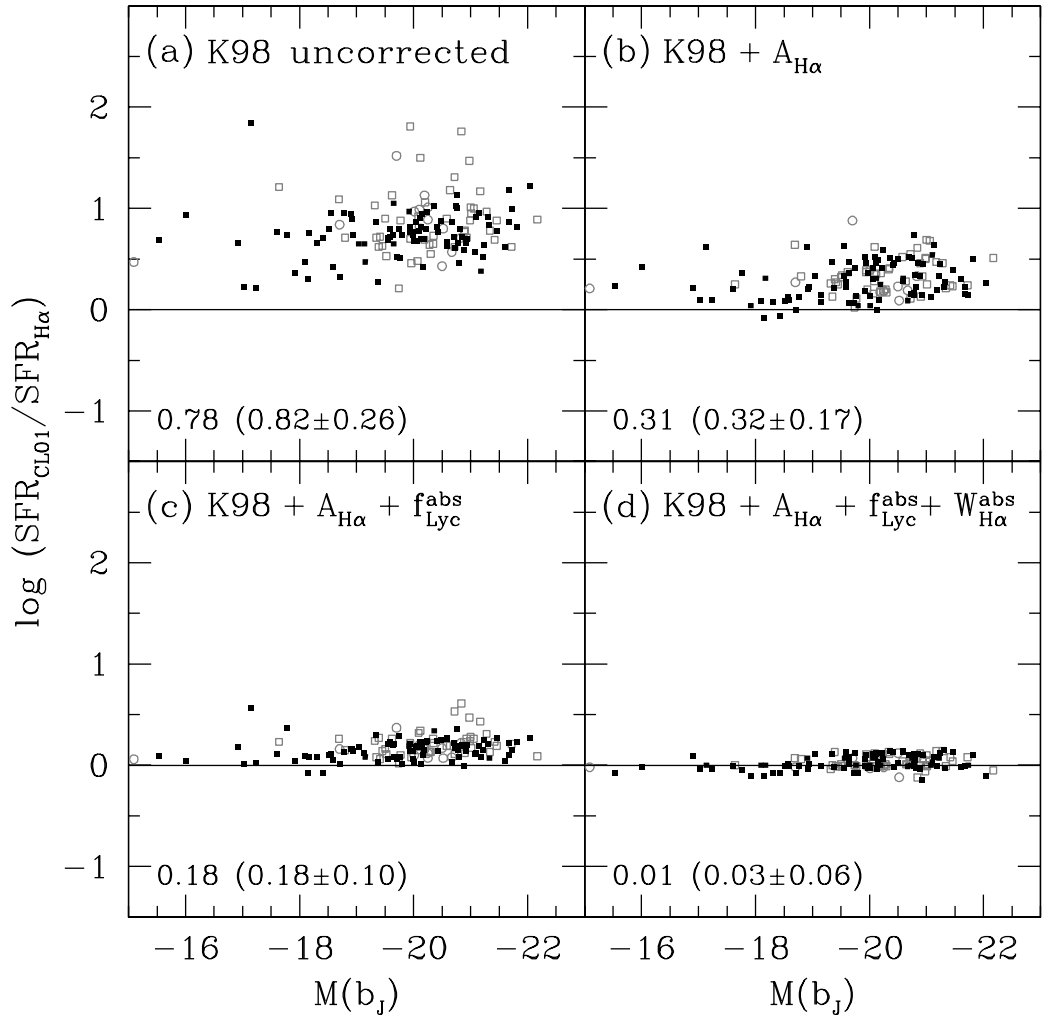}}
\caption{
Breakdown of the difference between the CL01 SFR estimates and \sfrha\ plotted
against absolute rest-frame \bj\ magnitude for the 149 SAPM galaxies with fluxes
and equivalent widths of \ha, \oii, and one or both of \nii\ and \sii, as well 
as IRAS 60 and 100$\,\mu$m fluxes.
Symbols are as described in Fig.~3. (a) Comparison of the CL01 SFRs with those 
derived from Equation (\ref{eqha}), {\em before} applying any correction for \ha\ 
attenuation. (b) After correcting the \ha-derived SFRs of individual galaxies using
the estimated attenuations shown in Fig.~5. (c) After further correction for the
fraction of ionizing photons absorbed by dust in \hii\ regions. (d) After further
correction for stellar absorption. In each panel, the median $\Delta\log(\sfr)$ 
offset for galaxies with $M(\bj)<-19$ is indicated along with the associated mean
offset and the rms scatter (in parentheses).}
\end{figure*}

\begin{figure*}
\epsfxsize=15cm
\hfill{\epsfbox{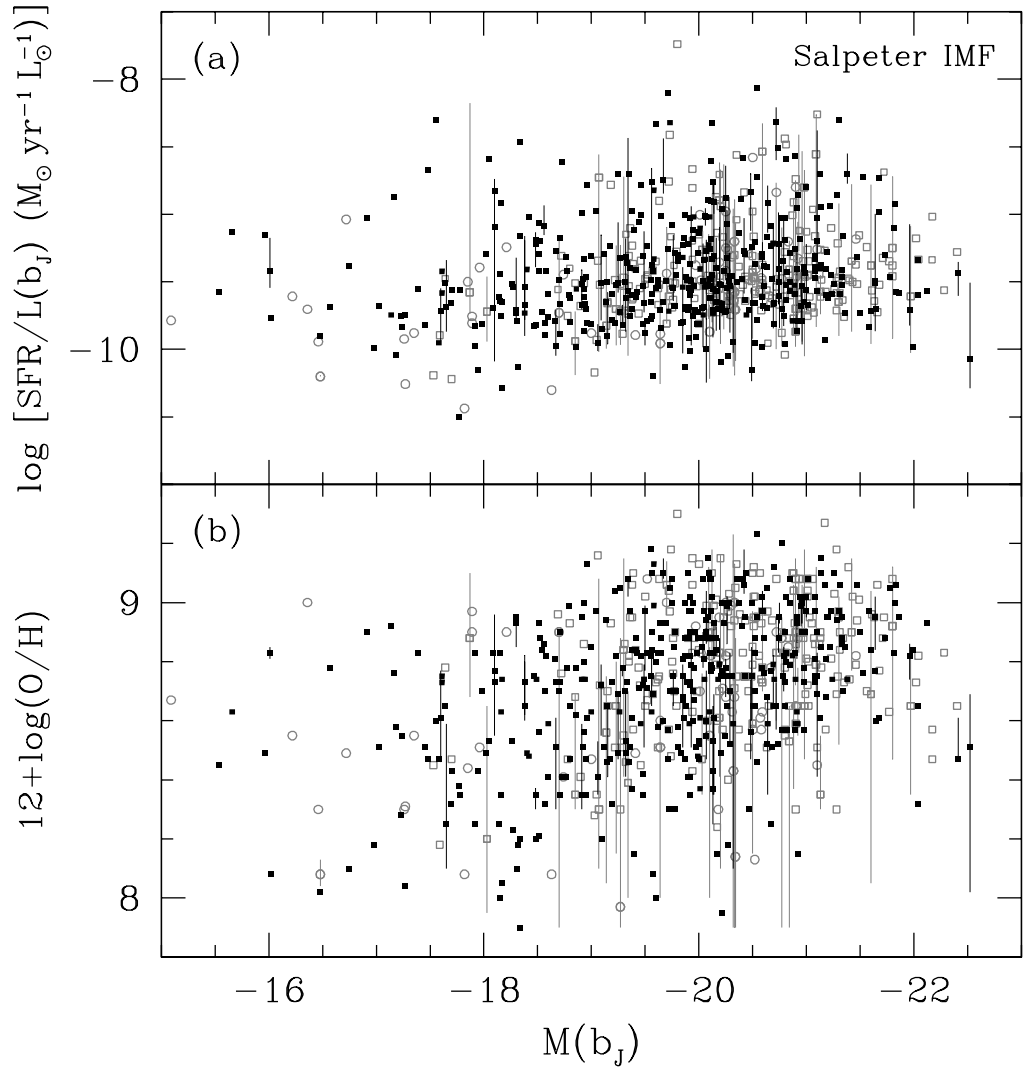}}
\caption{
Estimated SFRs and gas-phase oxygen abundances of the 705 SAPM galaxies with
fluxes and equivalent widths of \ha, \oii, and one or both of \nii\ and \sii\
plotted against absolute rest-frame \bj\ magnitude. Symbols are as described in 
Fig.~3. (a) Star formation rate per unit absolute \bj\ luminosity. (b) Gas-phase 
oxygen abundance [$12+\log (\oh)_\odot\approx8.8$]. For clarity, only 1 in 7 error
bars are shown to indicate the uncertainties in the SFRs and gas-phase oxygen
abundances derived using the CL01 models (as discussed in Section~4.2).}
\end{figure*}

\begin{figure*}
\epsfxsize=15cm
\hfill{\epsfbox{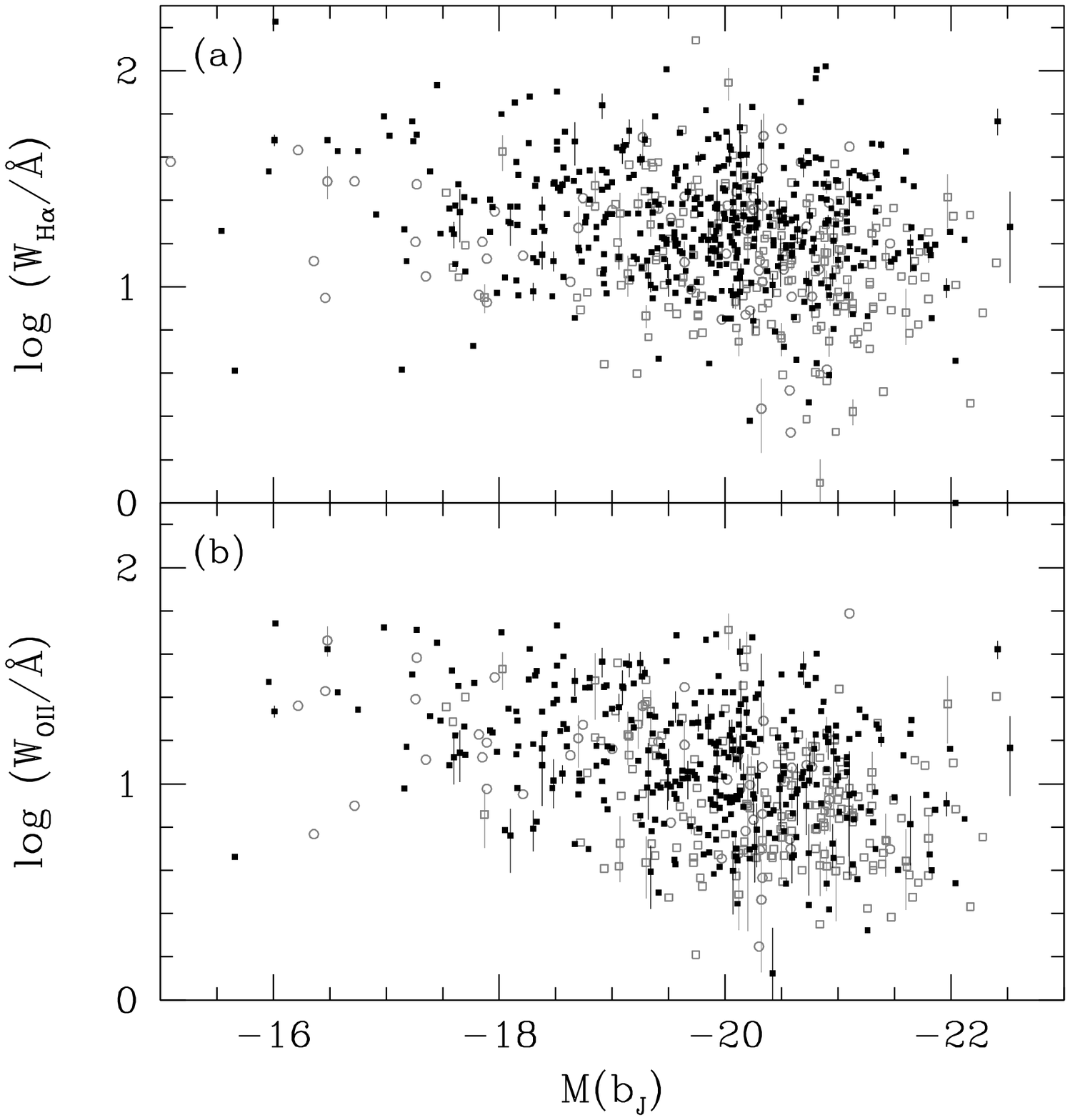}}
\caption{
\ha\ and \oii\ rest-frame emission equivalent widths of the 705 SAPM galaxies
with fluxes and equivalent widths of \ha, \oii, and one or both of \nii\ and \sii\
plotted against absolute rest-frame \bj\ magnitude. For clarity, only 
1 in 7 observational error bars are shown.}
\end{figure*}
\bsp

\label{lastpage}


\begin{thebibliography}{99}

\bibitem{b1} Baldi, A., Bardelli, S., Zucca, E., 2001, MNRAS, in press (astro-ph/0101416)
\bibitem{b1} Balogh, M.L., Schade, D., Morris, S.L., Yee, H.K.C., 
             Carlberg, R.G., Ellingson, E., 1998, ApJ, 504, L75
\bibitem{b1} Balogh, M.L., Christlein, D., Zabludoff, A., Zaritsky, D., 2001, ApJ, in press
             (astro-ph/0104042)
\bibitem{b1} Barton, E., Geller, M.J., Kenyon, S.J., 2000, ApJ, 530, 660                  
\bibitem{b1} Bell, E.F., Kennicutt, R.C., 2001, ApJ, 548, 681                               
\bibitem{b1} Blanton, M. , Lin, H., 2000, ApJ, 543, L125                                  
\bibitem{b1} Blitz, L. , Shu, F. H., 1980, ApJ, 238, 148
\bibitem{b1} Bromley, B.C., Press, W.H., Lin, H., Kishner, R.P., 1998, ApJ, 505, 25
\bibitem{b1} Bruzual A. G., Charlot S., 1993, ApJ, 405, 538
\bibitem{b1} Buat V., Xu C., 1996, A\&A, 306, 61
\bibitem{b1} Calzetti D., 1997, AJ, 113, 162
\bibitem{b1} Calzetti D., Kinney A. L., Storchi-Bergmann T., 1994,
             ApJ, 429, 582
\bibitem{b1} Carter,  B. J., Fabricant, D. G., Geller, M. J., Kurtz, M. J. 2001,
	     AJ, in press (astro-ph/0107258)
\bibitem{b1} Charlot S., Fall S. M., 2000, ApJ, 539, 718
\bibitem{b1} Charlot S., Longhetti, M., 2001, MNRAS, 323, 887
\bibitem{b1} Christlein, D., 2000, ApJ, 545, 145              
\bibitem{b1} Cowie, L.L., Songaila, A., Hu, E.M., Cohen, J.G., 1996, AJ, 112, 839          
\bibitem{b1} Cowie, L.L., Songaila, A., Barger, A.J., 1999, AJ, 118, 603          
\bibitem{b1} Degioia-Eastwood, K., 1992, ApJ, 397, 542
\bibitem{b1} Edmunds, M. G., Pagel, B. E. J. 1984, MNRAS, 211, 507
\bibitem{b1} Ellis, R.S., Colless, M., Broadhurst, T., Heyl, J., Glazebrook, K., 1996,
              MNRAS, 280, 235
\bibitem{b1} Fanelli M. N., O'Connell R. W., Thuan T. X., 1988,
             ApJ, 334, 665
\bibitem{b1} Ferguson A. M. N., Wyse R. F. G, Gallagher J. S.,
             Hunter D. A., 1996, AJ 111, 2265
\bibitem{b1} Ferland G., 1996, Hazy, A Brief Introduction to Cloudy. Internal
             Report, Univ. Kentucky, USA
\bibitem{b1} Folkes, S., Ronen, S., Price, I., Lahav, O., Colless, M., Maddox, S.,
	     Deeley, K., Glazebrook, K. et al, 1999, MNRAS, 308, 459
\bibitem{b1} Gallagher J. S., Hunter D. A., Bushouse H., 1989, 
             AJ, 97, 700
\bibitem{b1} Gallego, J., Zamorano, J., Aragon-Salamanca, A., Rego, M., 
             1995, ApJ, 455, L1 
\bibitem{b1} Hammer, F., Flores, H., Lilly, S.J., Crampton, D., Le Fevre, O., Rola, C.,
             Mallen-Ornelas, G., Schade, D., Tresse,L., 1997, ApJ, 481, 49
\bibitem{b1} Hashimoto, Y., Oemler. A., Lin, H., Tucker, D.L., 1998,
             ApJ, 499, 589
\bibitem{b1} Jansen R. A., Franx M., Fabricant D., 2001, ApJ, in press
	     (astro-ph/0012485)
\bibitem{b1} Heckman, T. M., Robert, C., Leitherer, C., Garnett, D. R.,
	     van der Rydt, F. 1998, ApJ, 503, 646
\bibitem{b1} Helou, G., Khan, I.R., Malek, L., Boehmer, L., 1988, ApJS, 68, 151
\bibitem{b1} Hoopes C. G., Walterbos R. A. M., Greenwalt B. E., 1996,
             AJ, 112, 1429
\bibitem{b1} Hunter D. A., Gallagher J. S, 1990, ApJ, 362, 480
\bibitem{b1} Kennicutt R. C., 1983a, ApJ, 272, 54
\bibitem{b1} Kennicutt R. C., 1983b, A\&A, 120, 219
\bibitem{b1} Kennicutt R. C., 1984, ApJ, 287, 116
\bibitem{b1} Kennicutt R. C., 1992a, ApJS, 79, 255
\bibitem{b1} Kennicutt R. C., 1992b, ApJ, 388, 310
\bibitem{b1} Kennicutt R. C., 1998, ARA\&A, 36, 189
\bibitem{b1} Kennicutt R. C., Tamblyn P., Congdon C. E., 1994,
             ApJ, 435, 22
\bibitem{b1} Kobulnicky H. A., Kennicutt, R.C., Pizagno J.L. 1999, ApJ, 514,
             544
\bibitem{b1} Kochanek, C. S., Pahre, M. A., Falco, E. E. 2001, ApJ, 
	     submitted (astro-ph/0011458)
\bibitem{b1} Leitherer C., Heckman T. M., 1995, ApJS, 96, 9
\bibitem{b1} Loveday J., Peterson B. A., Maddox S. J., Efstathiou G.,
	     1996, ApJS, 107, 201
\bibitem{b1} Loveday. J., Tresse, L., Maddox, S. J. 1999, MNRAS, 310, 281
\bibitem{b1} Madau, P., Ferguson, H.C., Dickinson, M.E., Giavalisco, M., 
             Steidel, C.C., Fruchter, A., 1996, MNRAS, 283, 1388
\bibitem{b1} Martin C. L., 1997, ApJ, 491, 561
\bibitem{b1} Mathis S. J., 1986, PASP, 98, 995
\bibitem{b1} Meurer G. R., Heckman T. M. Calzetti D., 1999, ApJ, 521, 64
\bibitem{b1} Misiriotis, A., Popescu, C.C., Tuffs, R., Kylafis, N.D., 2001,
             A\&A, in press (astro-ph/0104346)
\bibitem{b1} Niklas, S., Klein, U., Wielebinski, R., 1997, A\&A, 322, 19 
\bibitem{b1} Peterson B. A., Ellis R. S., Efstathiou G., Shanks T., 
	     Bean A. J., Fong R., Zen-Long Z., 1986, MNRAS, 221, 233
\bibitem{b1} Petrosian, V., Silk, J., Field, G. B. 1972, ApJ, 117, L69
\bibitem{b1} Poggianti, B. M., Smail, I., Dressler, A., Couch, W. J.,
	     Barger, A. J., Butcher, H., Ellis, R. S., Oemler, A. 1999,
	     ApJ, 518, 576
\bibitem{b1} Singleton C., 2001, PhD thesis, Univ. of Nottingham
\bibitem{b1} Sullivan, M., Treyer, M. A., Ellis, R. S., Bridges, T. J.,
	     Milliard, B., Donas, J., 2000, MNRAS, 312, 442
\bibitem{b1} Spitzer L., 1978, 
             Physical Processes in the Interstellar Medium.
             Wiley, New York, p. 333
\bibitem{b1} Tresse L., Maddox S.J., 1998, 495, 691
\bibitem{b1} Tresse L., Maddox S.J., Loveday J., Singleton C., 1999,
	     MNRAS, 310, 262
\bibitem{b1} York, D.G., Adelman, J., Anderson J.E., Anderson, S.F., Annis, J., 
             Bahcall, N.A., Bakken, J.A.,
             Barkhouser, R. et al, 2000, AJ, 120, 1579
\bibitem{b1} Zaritsky, D., Kennicutt, R. C., Huchra, J. P. 1994, ApJ, 420, 87
\bibitem{b1} Zucca, E., Zamorani, G., Vettolani, G, Cappi, A., Merighi, R.,
             Mignoli, M., Stirpe, G.M., MacGillivray, H. et al., 1997, A\&A, 326, 477
\end{thebibliography}
\end{document}